\documentclass[cits]{PoS}

\usepackage{multirow}
\usepackage{chemarrow}
\usepackage{amsmath}
\def\siii{^3 \hskip -0.025in S _1}
\def\diii{^3 \hskip -0.025in D _1}
 \usepackage{graphicx}
 \usepackage{subfigure}

\title{Two-Nucleon Systems in a Finite Volume}

\ShortTitle{NN-Systems in a Finite Volume}

\author{\speaker{Ra\'ul A. Brice\~no}\footnote{In collaboration with Z. Davoudi, T. Luu, and M. J. Savage} \\
        Jefferson Laboratory, 12000 Jefferson Avenue, Newport
  News, VA 23606, USA\\
        E-mail: \email{rbriceno@jlab.org}}

\abstract{I present the formalism and methodology for determining the nucleon-nucleon scattering parameters from the finite volume spectra obtained from lattice quantum chromodynamics calculations.  Using the recently derived energy quantization conditions
and the experimentally determined  scattering parameters,
the bound state spectra for finite volume systems with overlap with the $\siii$-$\diii$ channel are predicted for a range of volumes.
It is shown that the
extractions of the infinite-volume deuteron binding energy
and the low-energy scattering parameters, including  
the S-D mixing angle,
are possible 
from  Lattice QCD calculations of two-nucleon 
systems with boosts  of $|\mathbf{P}| \leq \frac{2\pi}{\rm L}\sqrt{3}$
in volumes with spatial extents $ {\rm L}$ satisfying $10~\rm{fm} \lesssim {\rm L} \lesssim 14~\rm{fm}$.
}

\FullConference{31st International Symposium on Lattice Field Theory - LATTICE 2013\\
		July 29 - August 3, 2013\\
		Mainz, Germany}

\begin{document}

\section{Introduction}

Lattice quantum chromodynamics (LQCD) calculations of the deuteron and its properties would serve as
theoretical milestones on the path towards determining quantities of importance in low-energy nuclear physics from quantum chromodynamics (QCD). LQCD calculations are necessarily performed in a finite Euclidean spacetime. Therefore, it is necessary to construct formalism that connects the finite-volume observables determined via LQCD to the infinite-volume quantities of interest. Although the Euclidean nature of the calculations imposes challenges on the determination of few-body scattering quantities away from the kinematic threshold in the infinite volume limit \cite{Maiani:1990ca}, the fact that these calculations are performed in a \textit{finite volume} (FV) allows for the extraction of scattering parameters from the spectrum through the \textit{L\"uscher} method \cite{Luscher:1986pf, Luscher:1990ux}. This method, which has been widely used to extract scattering phase shifts of two-hadron systems from LQCD (see for example Refs. \cite{Beane:2011sc,  Dudek:2012gj}), has been generalized to multi-coupled channel two-body systems with total spin $S\leq1/2$ \cite{Liu:2005kr, Bernard:2010fp, Hansen:2012tf, Briceno:2012yi,  Li:2012bi} as well as three-particle systems \cite{Polejaeva:2012ut, Briceno:2012rv}. 

I review the generalization of this formalism for two-nucleon systems with arbitrary parity, spin, isospin, angular momentum and center of mass motion first presented in Ref.~\cite{Briceno:2013lba}\footnote{The quantization condition for NN-systems was first presented in Ref.~\cite{Beane:2003da}. In this study the NN-system was constrained to have overlap with S-wave scattering channels only. The only previous attempt to address the complexity of the NN-system, including the spin, isospin and angular momentum degrees of freedom, is by N. Ishizuka \cite{Ishizuka:2009bx}, where the quantization conditions for energy eigenvalues of a two-nucleon system at rest in the positive and negative parity isosinglet channels were obtained for $J\leq 4$.}. This FV formalism and the experimentally determined scattering parameters~\cite{NIJMEGEN} are utilized to predict the spectra of the positive-parity isoscalar channel at the physical light quark masses. I discuss how the $\siii$-$\diii$ low energy scattering parameters along with the deuteron binding energy
can be simultaneously extracted from LQCD calculations performed in cubic
volumes with fields subject to 
periodic BCs (PBCs) in the spatial directions. The methodology reviewed here was first presented in Ref.~\cite{Briceno:2013bda}.

 \section{Deuteron and the Finite Volume Spectrum
\label{sec:DeutFV}
}

The energy eigenvalues of two nucleons in a cubic volume with PBCs are determined by
the S-matrix.
The following determinant condition, 
\begin{eqnarray}
\det~[\mathcal{M}^{-1}+\delta \mathcal{G}^V]=0 ,
\label{eq:QC}
\end{eqnarray}
provides the relation between the infinite volume
on-shell scattering amplitude 
$\mathcal{M}$ 
and the FV  CM energy of the NN system
below the inelastic threshold~\cite{Briceno:2013lba}. Since relativistic effects are expected to be suppressed for NN-systems, the discussion here will be restricted to nonrelativistic (NR) systems,
and as such the energy-momentum relation is 
${\rm E}_{NR}={\rm E}-2M={\rm E}_{NR}^* +\frac{\mathbf{P}^2}{4M}= \frac{k^{*2}}{M}+\frac{\mathbf{P}^2}{4M}$,
where $({\rm E}_{NR}, \mathbf{P})$ are the total NR energy and momentum of the
system, and $({\rm E}_{NR}^*,\mathbf{k}^*)$ are the NR CM energy and CM momentum. 
The subscript will be dropped for the remainder of the paper, simply denoting
${\rm E}_{NR}^{(*)}$ as ${\rm E}^{(*)}$. Due to the PBCs, the total momentum is  discretized, 
$\mathbf{P}=\frac{2\pi}{\rm L}\mathbf{d}$,
with $\mathbf{d}$ being an integer triplet referred to as the boost
vector and ${\rm L}$ the spatial extent of the lattice. 
$\delta \mathcal{G}^V$ can be written as a matrix in the $\left|JM_J(lS)\right
\rangle$-basis, 
where $J$ is the total angular momentum and $M_J$ is its azimuthal component, $l$ and $S$ are the orbital angular momentum and the total spin of the
channel, respectively. 
In this basis, the matrix elements of $\delta \mathcal{G}^V$ are,
\begin{eqnarray}
&& \left[\delta\mathcal{G}^V\right]_{JM_J,IM_I,lS;J'M_J',I'M_I',l'S'}=\frac{iMk^*}{4\pi}\delta_{II'}\delta_{M_IM_I'}\delta_{SS'}\left[\delta_{JJ'}\delta_{M_JM_J'}\delta_{ll'} +i\sum_{l'',m''}\frac{(4\pi)^{3/2}}{k^{*l+1}}c_{l''m''}^{\mathbf{P}}(k^{*2};{\rm L}) \right.
\nonumber\\
&&\hspace{-1cm} \qquad \qquad \qquad \qquad ~ \left .  \times \sum_{M_l,M_{l'},M_S}\langle JM_J|lM_l,SM_S\rangle \langle l'M_{l'},SM_S|J'M_J'\rangle \int d\Omega~Y^*_{ l,M_l}Y^*_{l'',m''}Y_{l',M_{l'}}\right],
\label{deltaG}
\end{eqnarray}
and are evaluated at the on-shell momentum of each nucleon in the CM frame, 
$k^*=\sqrt{M {\rm E}^*-{|\mathbf{P}|^2}/{4}}$. 
$\langle JM_J|lM_l SM_S\rangle$ and $\langle l'M_{l'} SM_S|J'M_J'\rangle$ are
Clebsch-Gordan coefficients, 
and $c_{lm}^{\mathbf{d}}(k^{*2}; {\rm L})$ is a kinematic function related to
the 
three-dimensional zeta function, 
$\mathcal{Z}^\mathbf{d}_{lm}$,~\cite{Luscher:1986pf, Luscher:1990ux, Rummukainen:1995vs, Christ:2005gi,Kim:2005gf},
\begin{eqnarray}
\hspace{1cm} 
c^\mathbf{d}_{lm}(k^{*2}; {\rm L})
=\frac{\sqrt{4\pi}}{\rm L^3}\left(\frac{2\pi}{\rm
    L}\right)^{l-2}\mathcal{Z}^\mathbf{d}_{lm}[1;(k^* {\rm L}/2\pi)^2],
\hspace{.5cm}
\mathcal{Z}^\mathbf{d}_{lm}[s;x^2]
= \sum_{\mathbf{n}}\frac{|\mathbf{r}|^lY_{l,m}(\mathbf{r})}{(r^2-x^2)^s},
\label{clm}
\end{eqnarray}
where $\mathbf{\mathbf{r}}=\mathbf{n}-\mathbf{d}/2$ with $\mathbf{n}$ an integer triplet.

Although the finite-volume matrix
$\delta\mathcal{G}^V$ is neither diagonal in the $J$
basis nor in the $lS$ basis, as is clear from the form of  Eq.~(\ref{deltaG}), for sufficiently low energies the determinant condition above, Eq.~(\ref{eq:QC}), can be truncated due to the kinematic suppression of the higher partial wave elements of the scattering amplitude  $\mathcal{M}$. By truncating the orbital angular momentum of the NN-system to $l\leq3$ and by considering three possible boost vector $\mathbf{d}=(0,0,0)$, $\mathbf{d}=(0,0,1)$ and $\mathbf{d}=(1,1,0)$, Ref.~\cite{Briceno:2013lba} provided the explicit form of the 49 independent quantization conditions (QCs) corresponding to the \emph{irreducible representations} (irreps) of the cubic ($O$), tetragonal ($D_4$) and orthorhombic ($D_2$) point groups. Furthermore, the QCs of the two irreps of the trigonal ($D_3$) group [corresponding to $\mathbf{d}=(1,1,1)$] that have overlap with the $\siii$-$\diii$ channel have been presented in Ref.~\cite{Briceno:2013bda}. 

Although these QCs will eventually be used to extract scattering parameters from the finite volume LQCD spectra, one can utilize them here in combination to the experimentally determined scattering parameters~\cite{NIJMEGEN} to predict the bound-state energies of the irreps that asymptote to the physical deuteron binding energy in the infinite-volume limit. To do this, it is convenient to parametrize the $J=1$ S-matrix using the Blatt-Biedenharn parameterization~\cite{Blatt:1952zza, PhysRev.93.1387}, 
\begin{eqnarray}
S_{(J=1)}=\left( \begin{array}{cc}
\cos\epsilon_1&-\sin\epsilon_1\\
\sin\epsilon_1&\cos\epsilon_1\\
\end{array} \right)
\left( \begin{array}{cc}
e^{2i\delta_{1\alpha}}&0\\
0&e^{2i\delta_{1\beta}}\\
\end{array} \right)
\left( \begin{array}{cc}
\cos\epsilon_1&\sin\epsilon_1\\
-\sin\epsilon_1&\cos\epsilon_1\\
\end{array} \right),
\label{eq:BBSmatrix}
\end{eqnarray}
where $\delta_{1\alpha}$ and $\delta_{1\beta}$ are the scattering phase shifts
corresponding to two eigenstates of the S-matrix; 
the so called ``$\alpha$'' and ``$\beta$'' waves respectively. $\epsilon_1$ is the mixing angle that parameterizes the coupling between the two partial waves. It is well known that at low energies and at the physical point, the $\alpha$-wave is predominantly S-wave with a small admixture of the D-wave, 
while the $\beta$-wave is predominantly D-wave with a small admixture of the
S-wave. The deuteron is an $\alpha$-wave bound state. 

\begin{figure}[ht!]
\begin{center}  
\subfigure[]{
\label{deut_cub-I}
\includegraphics[scale=0.18]{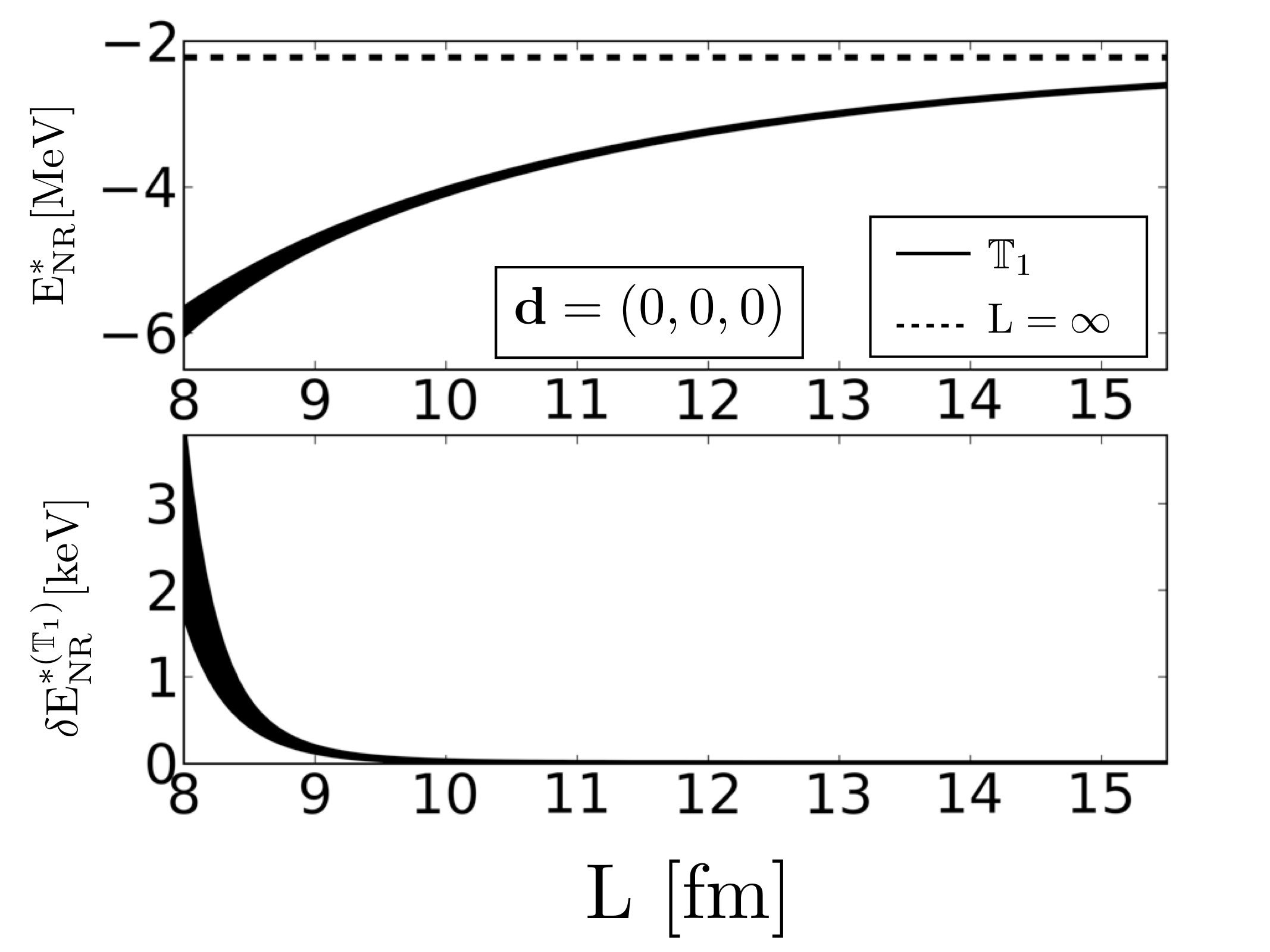}}
\subfigure[]{
\label{deut_trig}
\includegraphics[scale=0.18]{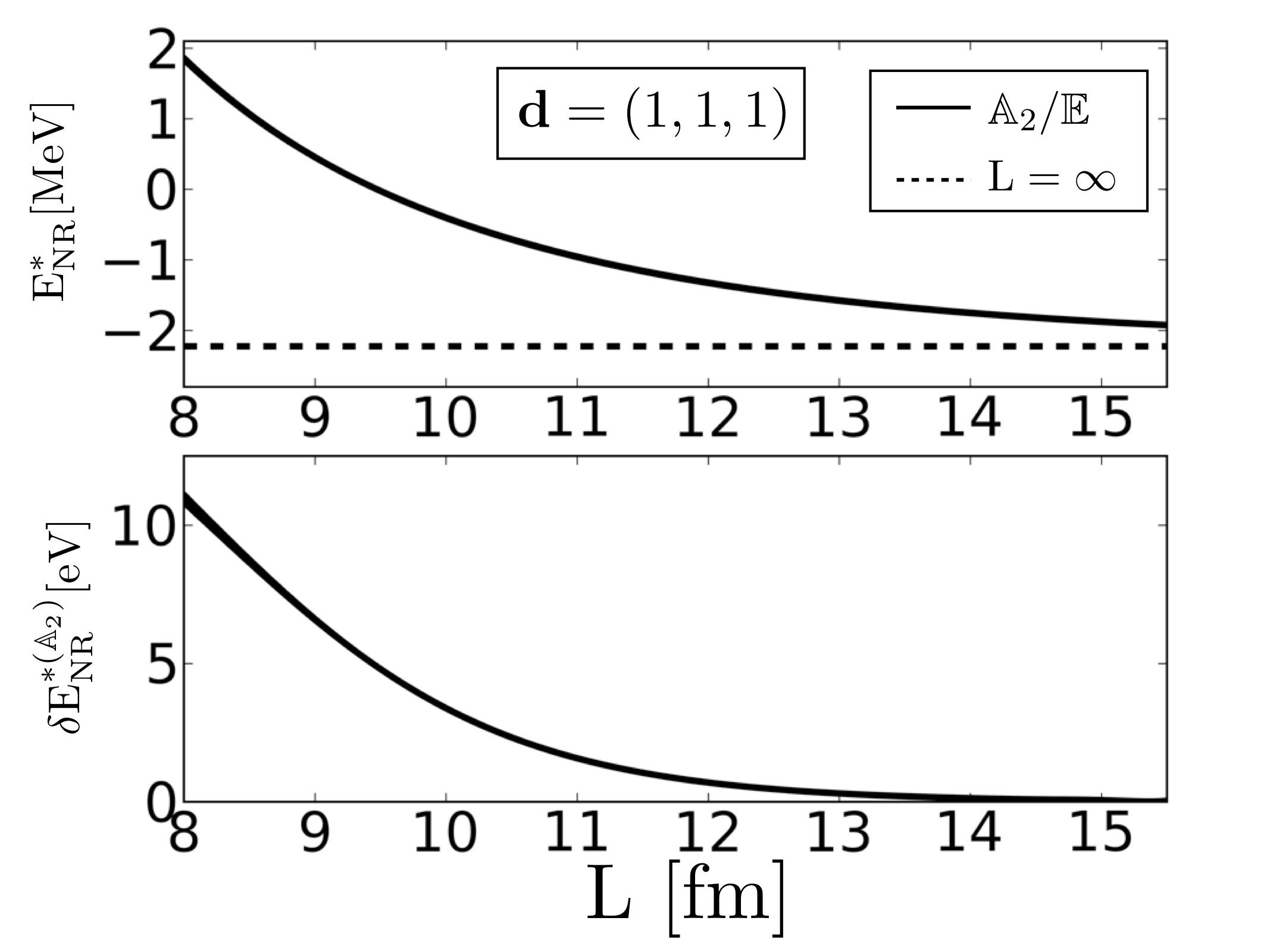}}
\caption[.]{(a) The upper panel shows the energy of two nucleons at rest in the positive-parity isoscalar channel  as a function of $\rm L$ extracted from the $\mathbb{T}_1$ QC, which can be deduced from Eq.~(\ref{eq:QC}) when the orbital momentum is truncated to $l\leq3$ and is explicitly written in Ref.~\cite{Briceno:2013lba}. The dashed line denotes the infinite-volume deuteron binding energy. The lower panel shows the contribution of the mixing angle to the  energy, $\delta {\rm E}^{*(\mathbb{T}_1)}= {\rm E}^{*(\mathbb{T}_1)}-{\rm E}^{*(\mathbb{T}_1)}(\epsilon_1=0)$. (b) The same quantities as in (a) for the NN system with ${\bf d}=(1,1,1)$ obtained from the  $\mathbb{A}_2/\mathbb{E}$ QCs.}
\label{deut_cub}
\end{center}
\end{figure}

\begin{figure}[ht!]
\begin{center}  
\subfigure[]{
\label{deut_cub-I}
\includegraphics[scale=0.18]{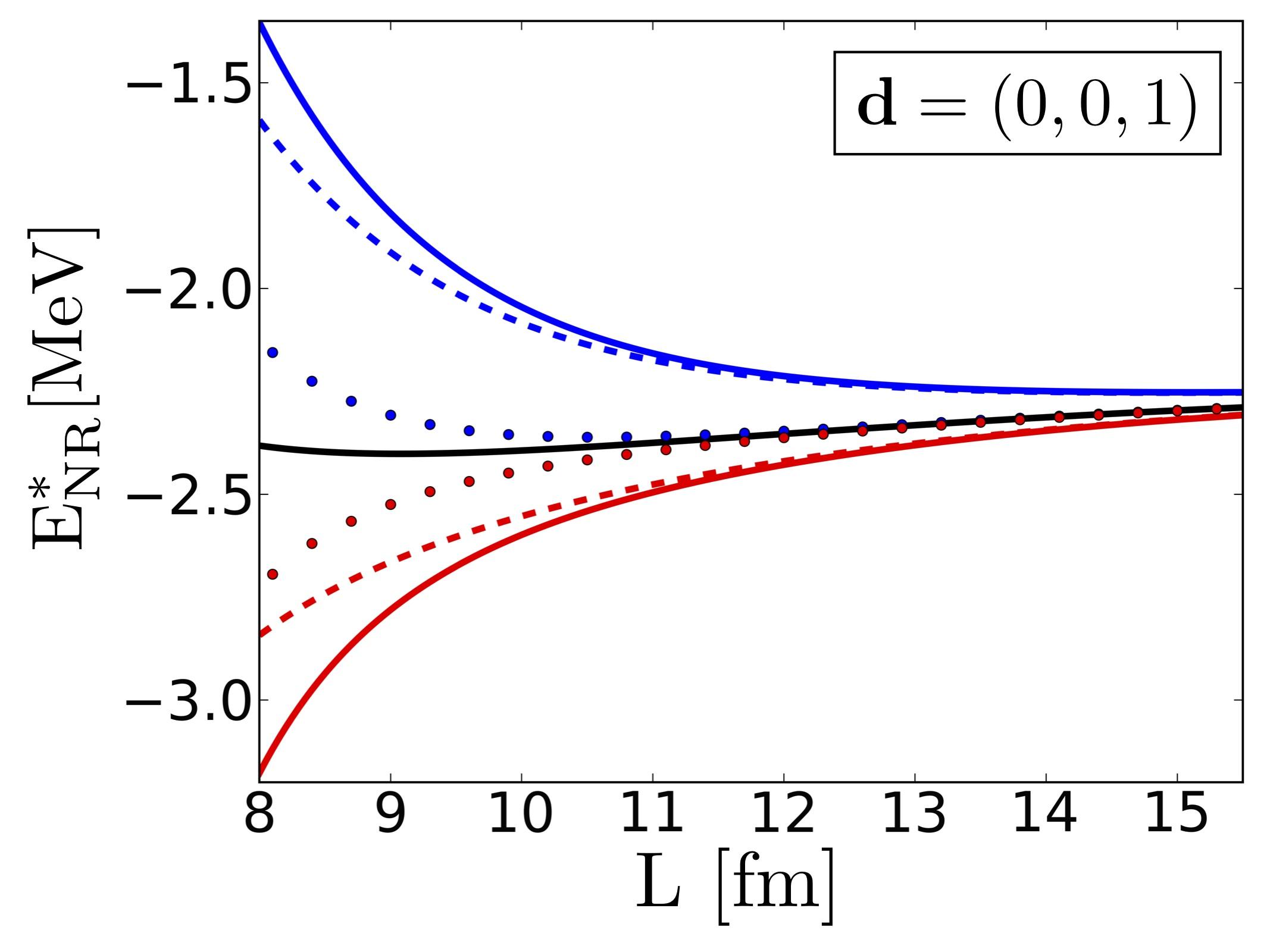}}
\subfigure[]{
\label{deut_trig}
\includegraphics[scale=0.18]{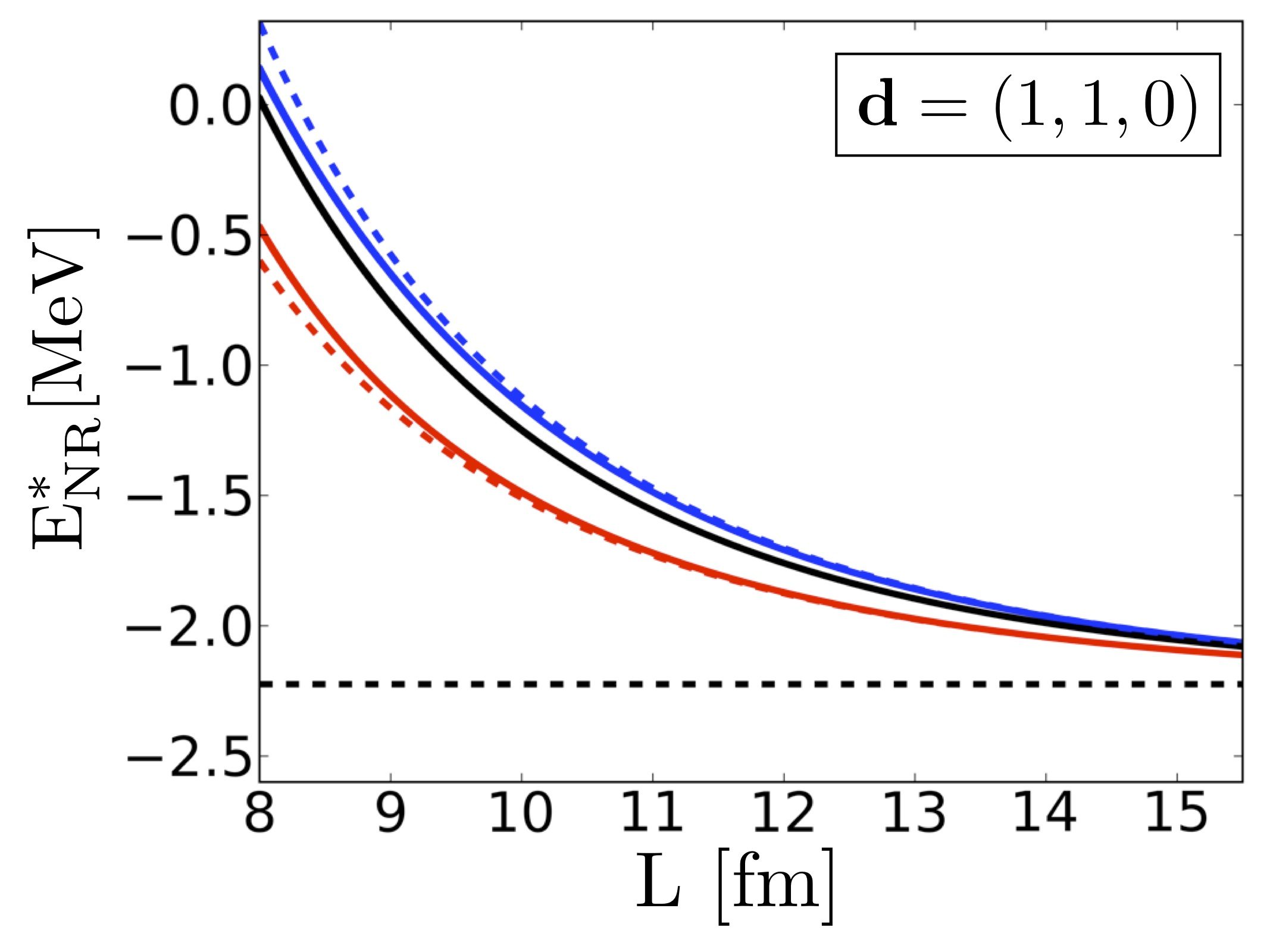}}
\caption[.]{(a) The energy of two nucleons in the
 positive-parity
isoscalar channel with $\mathbf{d}=(0,0,1)$ as a function of
  ${\rm L}$, 
extracted from the $\mathbb{A}_2$ (solid blue) and $\mathbb{E}$ (solid red) QCs. These QCs can be deduced from Eq.~(\ref{eq:QC}) when the orbital momentum is truncated to $l\leq3$ and are explicitly written in Ref.~\cite{Briceno:2013lba}. The dashed blue and red lines corresponds to the spectrum when $\delta_{1\beta}=\delta^{(^3D_2)}=\delta^{(^3D_3)}=0$  for the $\mathbb{A}_2$ and $\mathbb{E}$ irreps respectively. The dotted blue and red lines corresponds to the spectrum when $\epsilon_{1}=0$  for the $\mathbb{A}_2$ and $\mathbb{E}$ irreps respectively. The solid black line depicts the spectrum when all scattering parameters except the $\alpha$-wave phase shift are set to zero. 
(b) The same quantities as in (a) for the NN
system with ${\bf d}=(1,1,0)$ obtained from the  $\mathbb{B}_1$ (red) and $\mathbb{B}_2/\mathbb{B}_3$ (blue) QCs. The spectrum for $\epsilon_{1}=0$ is not shown, because it is indistinguishable from the black solid line. The black dashed line corresponds to the infinite volume deuteron binding energy. 
}
\label{deuttet}
\end{center}
\end{figure}

Figures~\ref{deut_cub} \& \ref{deuttet} show the spectra as a function of the volume  of the eight eigenstates that have overlap with the deuteron. An important observation is that, as can be observed from Fig.~\ref{deuttet}, the spectrum of a system with a boost vector $\mathbf{d}=(0,0,1)$ or $\mathbf{d}=(1,1,0)$ strongly depends on the mixing angle, $\epsilon_1$. This is certainly not the case for a system with a boost vector $\mathbf{d}=(0,0,0)$ or $\mathbf{d}=(1,1,1)$, as can be seen from Fig.~\ref{deut_cub}. 

\begin{figure}[t]
\begin{center} 
\subfigure[]{
\label{a_corr}
\includegraphics[scale=0.18]{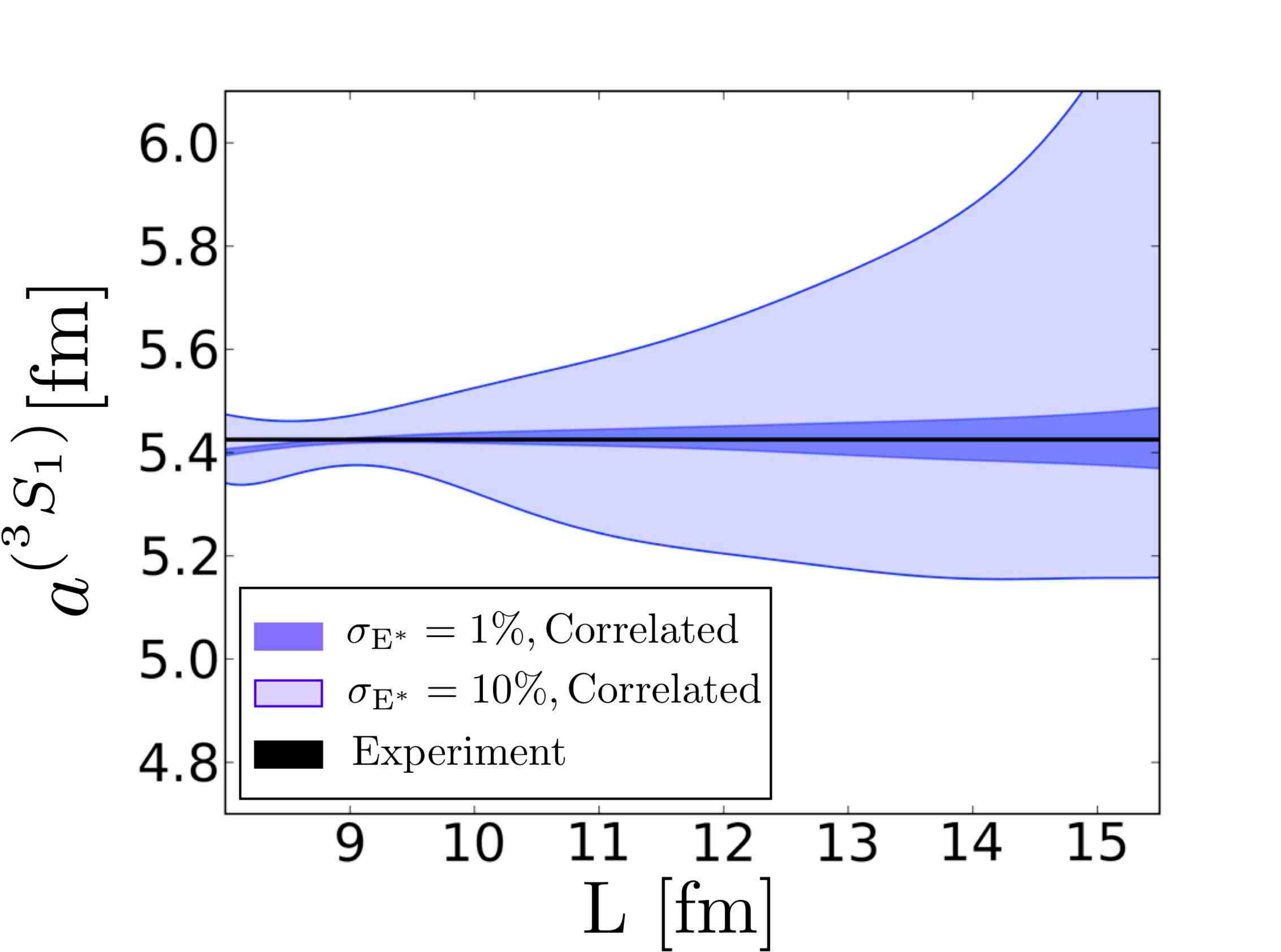}}
\subfigure[]{
\label{r_corr}
\includegraphics[scale=0.18]{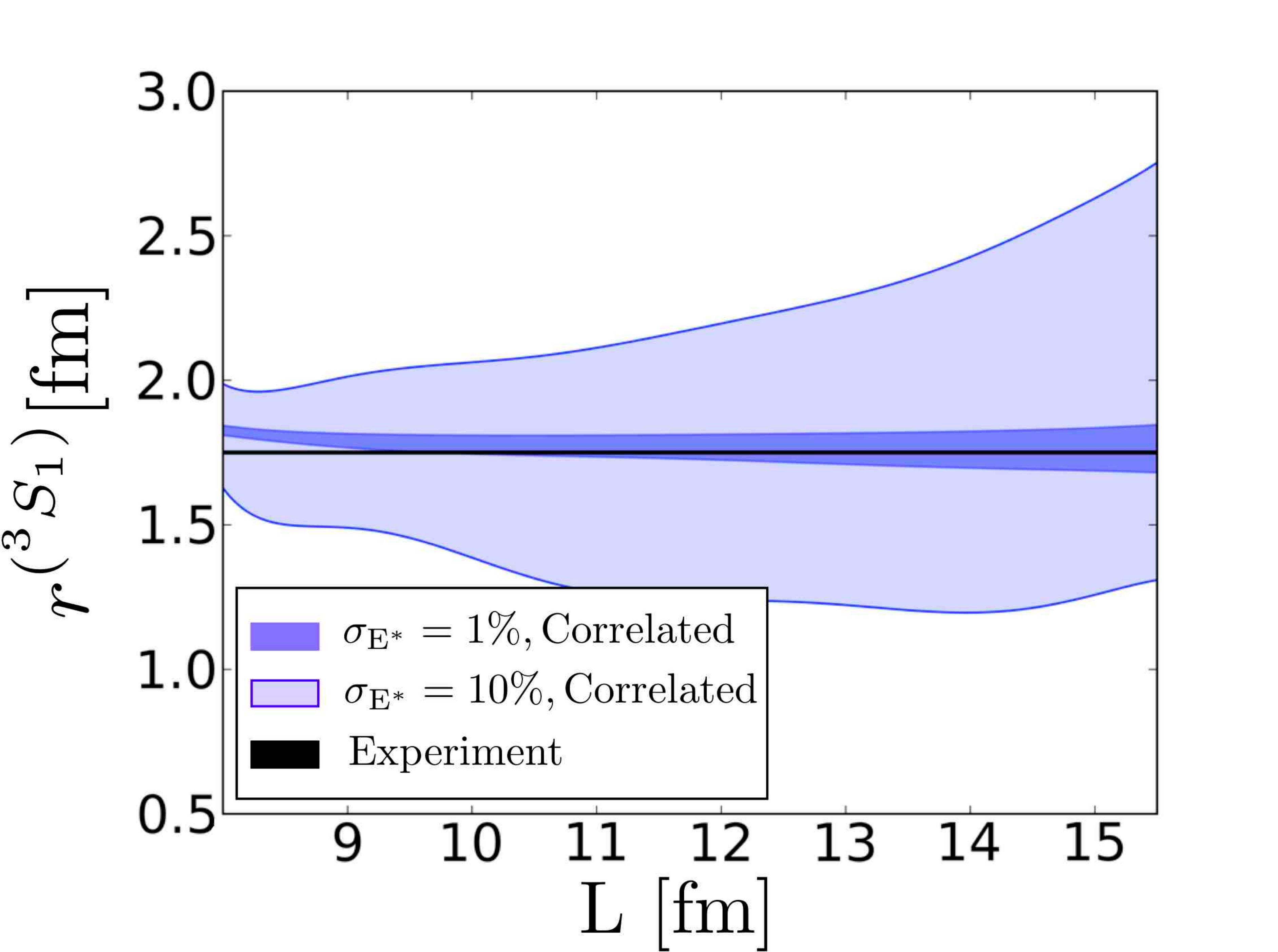}}
\subfigure[]{
\label{Bd_corr}
\includegraphics[scale=0.18]{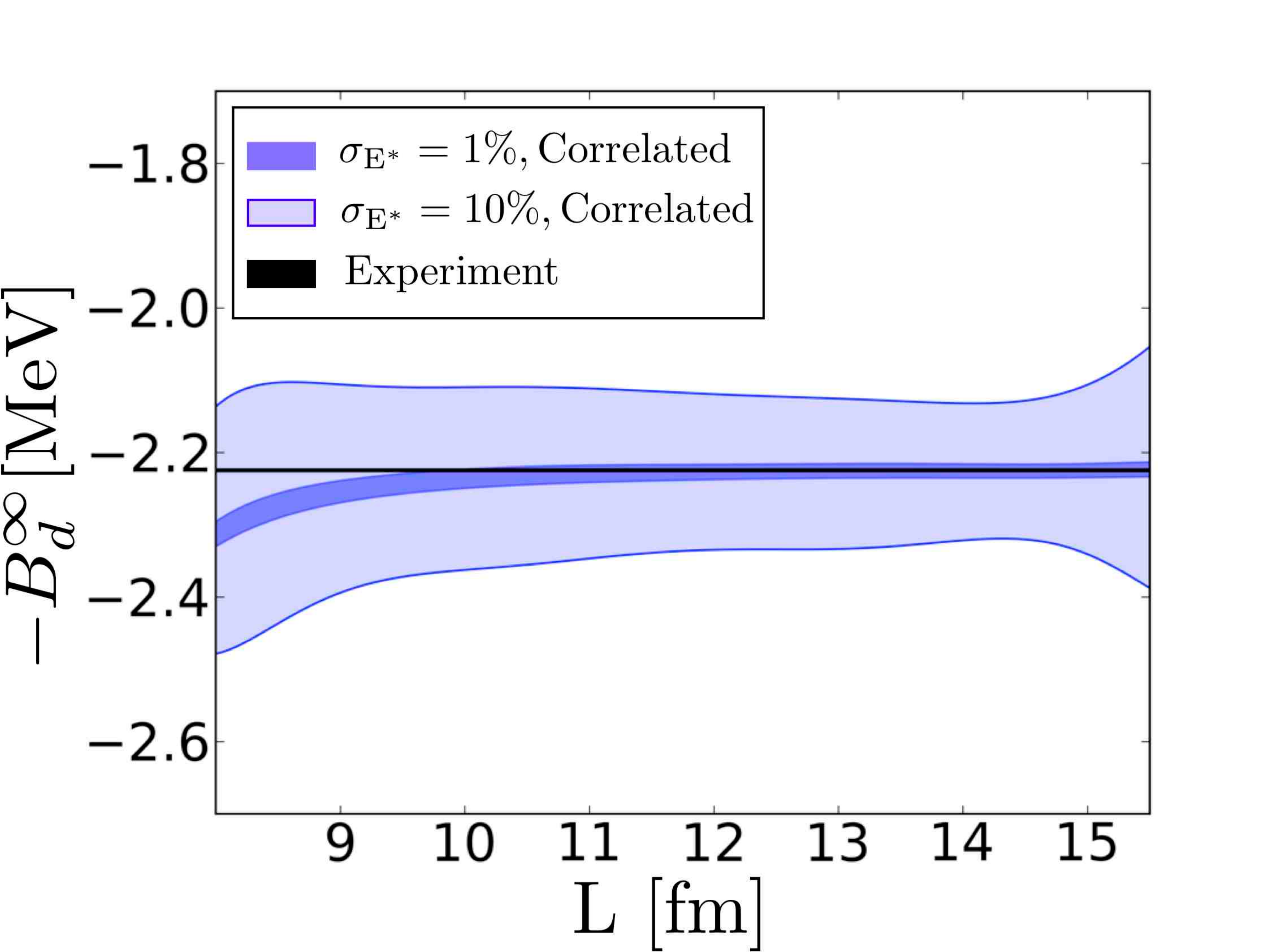}}
\subfigure[]{
\label{e1_corr}
\includegraphics[scale=0.18]{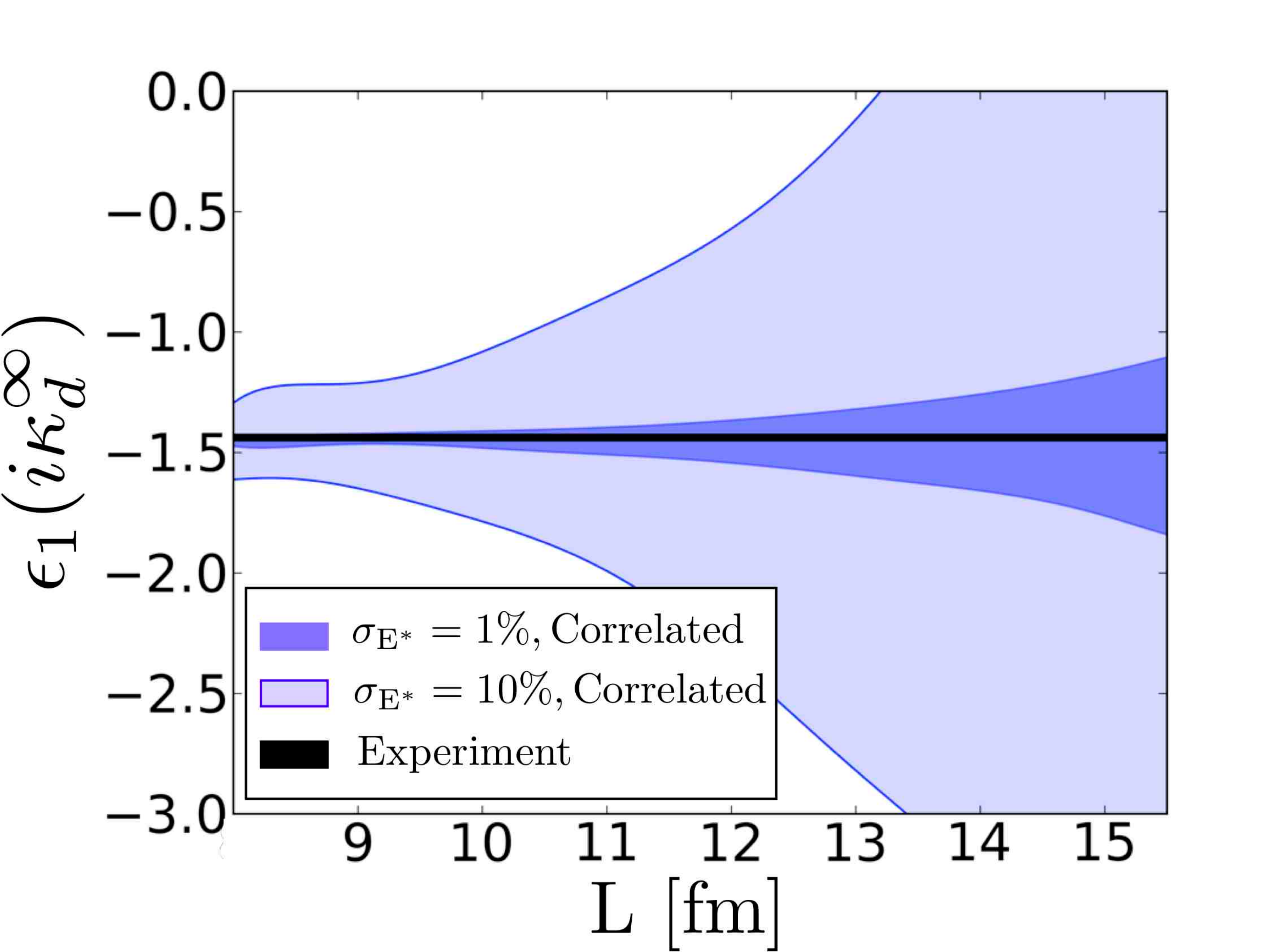}} 
\caption[.]{The values of $\{a^{(^3S_1)}, r^{(^3S_1)},B^{\infty}_d, \epsilon_1(i\kappa^\infty_d)\}$ obtained by fitting the six independent bound-state energies with $|\textbf{d}|\leq\sqrt{3}$ (depicted in Figs.~\ref{deut_cub} \& \ref{deuttet}) using the approximated QCs in Eqs. (\ref{appr-T1}-\ref{appr-EA2}), as discussed
in the text. 
The black lines denote
the experimental value of these quantities determined by fitting the scattering
parameters obtained from Ref.~\cite{NIJMEGEN}. 
The dark (light) inner (outer) band is the 
$1\sigma$ band corresponding to the energies being  determined with 1$\%$ (10$\%$) precision.  }
\label{fig:fakedata}
\end{center}
\end{figure}

The strong $\epsilon_1$-dependence of the of the bound states with ${\bf d}=(0,0,1)$ and $(1,1,0)$ compared 
to systems with $(0,0,0)$ and $(1,1,1)$ can be understood by careful investigation of the QCs given in Refs.~\cite{Briceno:2013lba, Briceno:2013bda}. It is straightforward to show that in the limit that the $\beta$-wave and $J\geq2$ D-wave phase shifts vanish, 
the QCs of the systems with $\mathbf{d}=(0,0,0)$ and  $\mathbf{d}=(1,1,1)$ reduce to a purely $\alpha$-wave condition
\begin{eqnarray}
\mathbb{T}_1&:&\hspace{.1cm}k^*\cot\delta_{1\alpha}-4 \pi
c_{00}^{(0,0,0)}(k^{*2}; {\rm L})=0 .
\label{appr-T1}\\
\mathbb{A}_2/\mathbb{E}&:&\hspace{.1cm}k^*\cot\delta_{1\alpha}-4 \pi
c_{00}^{(1,1,1)}(k^{*2}; {\rm L})=0 .
\label{appr-EA2}
\end{eqnarray}
In this limit the QCs for a systems with $\mathbf{d}=(0,0,1)$ and  $\mathbf{d}=(1,1,0)$ reduce to
\begin{eqnarray}
\mathbb{A}_2:\hspace{.1cm}
&& k^*\cot\delta_{1\alpha}
-4 \pi  c_{00}^{(0,0,1)}(k^{*2}; {\rm L})
\ =\ 
- \frac{4\pi}{\sqrt{5}k^{*2}}\ 
c_{20}^{(0,0,1)}(k^{*2}; {\rm L})
\ 
(\sqrt{2}s_{2\epsilon_1}-s_{\epsilon_1}^2)
,
\\
\mathbb{E}:\hspace{.1cm}
&& k^* \cot \delta_{1\alpha}
-4 \pi  c_{00}^{(0,0,1)}(k^{*2}; {\rm L})
\ =\  \frac{4\pi}{2\sqrt{5}k^{*2}}\ 
c_{20}^{(0,0,1)}(k^{*2}; {\rm L})
\ 
(\sqrt{2}s_{2\epsilon_1}-s_{\epsilon_1}^2)
,
\label{appr-E}
\end{eqnarray}
\begin{eqnarray}
\mathbb{B}_1:
\hspace{.1cm}
&& k^*\cot\delta_{1\alpha}
-4 \pi  c_{00}^{(1,1,0)}(k^{*2}; {\rm L})
\ =\ 
-
\frac{4\pi}{\sqrt{5}k^{*2}}\ 
c_{20}^{(1,1,0)}(k^{*2}; {\rm L})
\ 
(\sqrt{2}s_{2\epsilon_1}-s_{\epsilon_1}^2) ,
\\
\mathbb{B}_2/\mathbb{B}_3:
\hspace{.1cm}
&& k^* \cot \delta_{1\alpha}
-4 \pi  c_{00}^{(1,1,0)}(k^{*2}; {\rm L})
\ =\  
\frac{4\pi}{2\sqrt{5}k^{*2}}\ 
c_{20}^{(1,1,0)}(k^{*2}; {\rm L})
\ 
(\sqrt{2}s_{2\epsilon_1}-s_{\epsilon_1}^2),
\label{appr-B2}
\end{eqnarray}
where $s_{\epsilon_1}=\sin\epsilon_1$ and $s_{2\epsilon_1}=\sin2\epsilon_1$. Therefore, these QCs include corrections to the $\alpha$-wave limit that scale with $\sin \epsilon_1$ at LO. 
This explains the large deviations of the energy eigenvalues for these irreps from the purely S-wave values.  It is straightforward to show that corrections to these QCs scale $\sim \frac{1}{\rm L}e^{-2\kappa {\rm L}}\tan{\delta_{1\beta}}$ and 
$\frac{1}{\rm L}e^{-2\kappa {\rm L}}\tan{\delta_{D_{J=2,3}}}$, where $\kappa=\sqrt{B^{\rm{L}}_d M}=-ik^*$ is the FV bound state momentum and $B^{\rm{L}}_d$ denotes the FV binding energy.
 
In order to quantify the level of accuracy and precision with which the deuteron binding energy, $\alpha$-wave scattering length and effective range, and the mixing angle can be determined at low energies, I use the approximated QCs above, Eqs.~(\ref{appr-T1}-\ref{appr-B2}) along with the effective range expansion up to $\mathcal{O}(k^{*2})$ to simultaneously fit the six independent bound states energies with $|\textbf{d}|\leq\sqrt{3}$ predicted by the full QC, Eq.~\ref{eq:QC}, by truncating the orbital angular momentum to be $l\leq3$. This is done for NR energies that are assumed to be determined at the 1\% and 10\% level of precision. Also this is done for both correlated and uncorrelated energies. In Fig.~\ref{fig:fakedata} only the correlated results are presented, but in Ref.~\cite{Briceno:2013lba} it can be seen that these do not observably change when the energies are assumed to uncorrelated. Figure~\ref{fig:fakedata} allows us to conclude that with the approximated QCs above, one can in fact reliably determined the infinite volume deuteron binding energy and mixing angle from the determination of the bound state spectra at a single moderate volume satisfying $\kappa {\rm L} =2-2.5$. By determining the spectra at 1~\% at a single volume ${\rm L}\sim10~{\rm fm}$ one can accurately reproduce all of these quantities at the $\sim1\%$ level of precision.

\section*{Acknowledgments} 
The work reviewed in this talk was supported in part by the DOE grant No. DE-FG02-97ER41014.
RB would like to thank his collaborators Z. Davoudi, T. Luu, and M. J. Savage for their contributions to the work reviewed in this talk and many useful discussions.


\begin{thebibliography}{99}


\bibitem{Maiani:1990ca}
L.~Maiani and M.~Testa  
\href{http://www.sciencedirect.com/science/article/pii/0370269390906953}{{\em Phys. Lett.\/} {\bf B245} 585--590 (1990)}.

\bibitem{Luscher:1986pf}
M.~Luscher, \href{http://link.springer.com/article/10.1007\%2FBF01211097} {{\em Commun. Math. Phys.\/} {\bf 105} 153--188 (1986).}

\bibitem{Luscher:1990ux}
M.~Luscher,  \href{http://www.sciencedirect.com/science/article/pii/0550321391903666}{{\em Nucl. Phys.\/} {\bf B354} 531--578 (1991).}

\bibitem{Beane:2011sc}
S.~R.~Beane  {\em et~al.\/} (NPLQCD Collaboration)
\href{http://prd.aps.org/abstract/PRD/v85/i3/e034505} {{\em Phys. Rev. \/} {\bf
  D85} 034505 (2012),} \href{http://arxiv.org/abs/arXiv:1107.5023}{arXiv:1107.5023 [hep-lat].}

\bibitem{Dudek:2012gj}
  J.~J.~Dudek, R.~G.~Edwards and C.~E.~Thomas,
 \href{http://prd.aps.org/abstract/PRD/v86/i3/e034031}{{\em Phys. Rev.  \/} {\bf D86} 034031(2012)}, \href{http://arxiv.org/abs/arXiv:1203.6041}{arXiv:1203.6041 [hep-ph].} 

\bibitem{Liu:2005kr}
 C.~Liu, X.~Feng and S.~ He \href{http://www.worldscientific.com/doi/abs/10.1142/S0217751X06032150}{{\em Int. J. Mod. Phys.\/} {\bf A21} 847--850 (2006)}, \href{http://arxiv.org/abs/hep-lat/0508022}{hep-lat/0508022.}

\bibitem{Bernard:2010fp}
V.~Bernard , M.~Lage, U.~G.~Meissner and A.~Rusetsky \href{http://link.springer.com/article/10.1007\%2FJHEP01\%282011\%29019}{{\em JHEP\/} {\bf 1101} 019 (2011)}, \href{http://arxiv.org/abs/arXiv:1010.6018}{arXiv:1010.6018 [hep-lat].}

\bibitem{Hansen:2012tf}
M.~T.~Hansen  and S.~R.~Sharpe \href{http://prd.aps.org/abstract/PRD/v86/i1/e016007}{{\em Phys. Rev.\/} {\bf D86} 016007 (2012)}, \href{http://arxiv.org/abs/arXiv:1204.0826}{arXiv:1204.0826 [hep-lat].}

\bibitem{Briceno:2012yi}
R.~A.~Briceno and Z.~Davoudi (2012), \href{http://arxiv.org/abs/arXiv:1204.1110}{arXiv:1204.1110 [hep-lat].}

\bibitem{Li:2012bi}
N.~Li and C.~Liu  \href{http://prd.aps.org/abstract/PRD/v87/i1/e014502}{{\em Phys. Rev. \/} {\bf D87} 014502 (2013)}, \href{http://arxiv.org/abs/arXiv:1209.2201}{arXiv:1209.2201 [hep-lat].}

\bibitem{Polejaeva:2012ut}
K. Polejaeva and A.~Rusetsky \href{http://link.springer.com/article/10.1140\%2Fepja\%2Fi2012-12067-8}{{\em Eur. Phys. J. \/} {\bf A48} 67 (2012)}, \href{http://arxiv.org/abs/arXiv:1203.1241}{arXiv:1203.1241 [hep-lat].}

\bibitem{Briceno:2012rv}
R.~A.~Briceno and Z.~Davoudi \href{http://prd.aps.org/abstract/PRD/v87/i9/e094507}{{\em Phys. Rev. \/} {\bf D87} 094507 (2012)}, \href{http://arxiv.org/abs/arXiv:1212.3398}{arXiv:1212.3398 [hep-lat].}

\bibitem{Briceno:2013lba}
R.~A.~Briceno, Z.~Davoudi and T.~C.~Luu \href{http://prd.aps.org/abstract/PRD/v88/i3/e034502}{{\em Phys. Rev. \/} {\bf D88} 034502 (2013)}, \href{http://arxiv.org/abs/arXiv:1305.4903}{arXiv:1305.4903 [hep-lat].} 
\bibitem{Beane:2003da}
S.~R.~Beane, P.~Bedaque, A.~Parreno and M.~Savage \href{http://www.sciencedirect.com/science/article/pii/S0370269304002618}{{\em Phys. Lett. \/} {\bf B585}
  106--114 (2004)}, \href{http://arxiv.org/abs/hep-lat/0312004}{hep-lat/0312004}.

\bibitem{Ishizuka:2009bx}
N.~Ishizuka {\em PoS\/} {\bf LAT2009} 119 (2009) \href{http://arxiv.org/abs/arXiv:0910.2772}{arXiv:0910.2772 [hep-lat]}.

\bibitem{NIJMEGEN}
  Nijmegen Phase Shift Analysis, \href{http://nn-online.org/}{http://nn-online.org/}.

\bibitem{Briceno:2013bda}
R.~A.~Briceno, Z.~Davoudi, T.~C.~Luu and M.~J.~Savage (2013), \href{http://arxiv.org/abs/arXiv:1309.3556}{arXiv:1309.3556 [hep-lat]}.  

\bibitem{Rummukainen:1995vs}
K.~Rummukainen and S.~A.~Gottlieb \href{http://www.sciencedirect.com/science/article/pii/055032139500313H}{{\em Nucl. Phys. \/} {\bf B450} 397--436 (1995)}, \href{http://arxiv.org/abs/hep-lat/9503028}{hep-lat/9503028.}

\bibitem{Christ:2005gi}
N.~H.~Christ, C.~Kim and T.~Yamazaki \href{http://prd.aps.org/abstract/PRD/v72/i11/e114506}{{\em Phys. Rev. \/} {\bf D72} 114506 (2005)}, \href{http://arxiv.org/abs/hep-lat/0507009}{hep-lat/0507009.}

\bibitem{Kim:2005gf}
C.~Kim, C.~Sachrajda and S.~R.~Sharpe \href{http://www.sciencedirect.com/science/article/pii/S0550321305007133}{{\em Nucl. Phys. \/} {\bf B727} 218--243 (2005)}, \href{http://arxiv.org/abs/hep-lat/0507006}{hep-lat/0507006}.

\bibitem{Blatt:1952zza}
J.~M.~Blatt and L.~Biedenharn \href{http://prola.aps.org/abstract/PR/v86/i3/p399_1}{{\em Phys. Rev. \/} {\bf 86} 399--404 (1952).}

\bibitem{PhysRev.93.1387}
L.~C.~Biedenharn and J.~M.~Blatt \href{http://prola.aps.org/abstract/PR/v93/i6/p1387_1}{{\em Phys. Rev. \/} {\bf 93}(6) 1387--1394 (1954).} 

\end{thebibliography}
\end{document}